\documentclass[twocolumn,twoside,slac_two]{revtex4}
\usepackage{graphicx}
\usepackage{fancyhdr}
\begin{document}

\title{$\pi \pi$ scattering length measurements }

%

\author{S. Giudici  (on behalf of NA48 coll.) }
\affiliation{University of Pisa and INFN, Italy}

\begin{abstract}
The high statistics sample of $K^{+} \rightarrow \pi^+ \pi^0 \pi^0$ and
$K_{e4}$ decays collected by NA48/2 experiment at the CERN/SPS,
combined with good performances of the detector, allowed two independent precise measurements
of the  pion-pion scattering lengths. Principles of the measurement, 
methods and results will be reviewed.
\end{abstract}

\maketitle

\thispagestyle{fancy}


\section{Introduction}
In the low energy regime  $\pi \pi$ scattering cross section is dominated 
by the S-wave contribution and its effect. The action of scattering Matrix is just a rephasing of the two-pion state
$$S|\pi \pi > = e^{2i\delta_I} |\pi \pi>  ~~~~~~ I=0,2$$  
where $I=0,2$ represents the only two allowed Isospin states because of Bose-statistics. 
The two phases may be expressed as $\delta_I = a_I k$
where $k$ is the pion momentum in the center of mass frame and parameters $a_0 $ and $a_2$ 
are called S-wave pion scattering lengths and are fundamental quantities of Chiral Perturbation Theory.  
Experimentally, one may access to the two parameters by studying the 
$\pi \pi$ final state rescattering effects which are related to $a_0$ and $a_2$. In the following we discuss 
separately $K\rightarrow 3\pi$ and $K_{e4}$ cases.  
\begin{figure}[h]
\centering
\includegraphics[width=40mm]{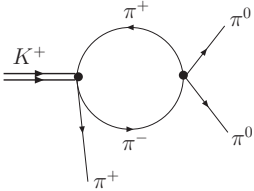}
\caption{1 loop Feynmann Diagram for $K\rightarrow \pi^+ \pi^0  \pi^0$} \label{fey}
\end{figure}

\begin{figure}[h]
\centering
\includegraphics[width=50mm]{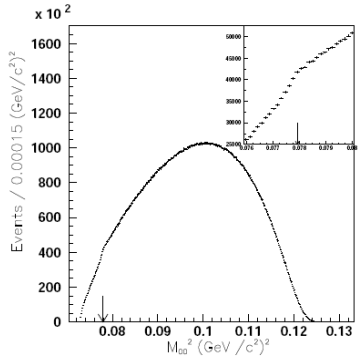}
\caption{Distrubution of the two neutral pion invariant mass squared $m_{00}$ and zoom of the Cusp.}  \label{cusp}
\end{figure}

\section{$K\rightarrow 3\pi$}
Let's consider the contribution of the 1 loop charge exchange Feynmann diagram, shown in Fig. \ref{fey},   
to the decay $K\rightarrow \pi^+ \pi^0  \pi^0$. If $m_+$ indicates the charged pion mass, one may expect a threshold 
effect when the two neutral pion invariant mass takes the value  $m_{00} = 2m_+$ which corresponds to a square root like singularity 
in the amplitude. The interference of this particular diagram with the tree level process generates a discontinuity
in the first derivative of the $m_{00}$ spectrum. This singular behaviour of cross section near threshold 
is a well known phenomenon studied in the past by Wigner ~\cite{wigner} and is commonly called ``Wigner Cusp'' or simply
``Cusp'' in the following.  The cusp structure has been clearly seen by experiment NA48 with a sample of $60 \times 10^{6} $ fully 
reconstructed  $K\rightarrow \pi^+ \pi^0  \pi^0$ events. The experimental distribution of $m_{00}$ is shown in Fig.
(\ref{cusp}).  A calculation of the effect has been published by N. Cabibbo and G. Isidori \cite {cabibbo} \cite{cabibbo-isidori} where  
the amplitude of the dominant charge-exchange diagram is parametrized in terms of pion scattering lengths as 
$$ {\cal M } \propto  (a_0-a_2) \sqrt{1-  \left ( \frac{m_{00}}{2m_+} \right )^2 }  $$
Higher order terms involve different linear combinations of $a_0$ and $a_2$.  
The cited theoretical models have been implemented by NA48  in a fitting procedure to the experimental data in order to extract the 
scattering lengths parameters. The analysis is discussed in reference \cite{cuspexp}.
The fitting region has been restricted to the range $0.074 \le m^2_{00} \le 0.097 (\mbox{GeV/c}^2)^2$ because the theoretical model
adopted was developed as a Taylor expansion around the cusp threshold. A few bins around the cusp have been also excluded 
from the fitting to limit the effect of Coulomb corrections and Pionium formation (i.e. $\pi^+ \pi^-$ electromagnetic
bound state) \cite{Silagazde}.
The quality of the fit is good  ($\chi ^2 = 145.5/139$) and the values found are:
\begin{eqnarray}
&(a_0 - a_2)m_+  = 0.268 \pm 0.010 (stat) \nonumber \\  
&\pm 0.004 (syst) \pm 0.013 (ext) \nonumber \\    
& a_2 m_+  = -0.041 \pm  0.022 (stat) \pm 0.014 (syst) \nonumber
\end{eqnarray}

The two statistical errors from the fit are strongly correlated with a correlation
coefficient of $-0.858$. The external error quoted reflects the impact of uncertainties on   
branching ratios and form factors entering in the fitting procedure as well 
as an additional $\pm 5 \%$ theoretical error estimated in reference \cite {cabibbo-isidori} 
quoted as the uncertainty  from neglecting higher-order terms and radiative 
corrections in the rescattering model.

A theoretical approach, alternative to Cabibbo-Isidori model, has been 
published in \cite{gasser}. In the future, this new model will be tested also on the sample 
of $\sim 100 \times 10^6$ $K_L \rightarrow 3\pi^0$, collected by NA48 in the 
year 2000, where a similar 
cusp structure, though attenuated with respect to the charged Kaon case, has been seen.  
Data are being re-analyzed. KTEV has recently published a study on the 
$K_L \rightarrow 3\pi^0$ Dalitz plot \cite{ktev} and the results found are in agreement with 
the NA48 measurements discussed here.
Other experiments have tried to look for an equivalent cusp structure in $\eta \rightarrow 3\pi^0$ 
but the statistical power reached so far is not enough \cite{wasa}.

\section{$K ^{\pm} \rightarrow \pi^+ \pi^- e^{\pm} \nu $}

\begin{figure}[h]
\centering
\includegraphics[width=40mm]{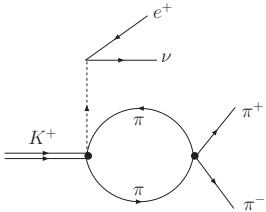}
\caption{Feynmann diagram for $Ke4$ decay with final state rescattering} \label{ke4fey}
\end{figure}

\begin{figure}[h]
\centering
\includegraphics[width=70mm]{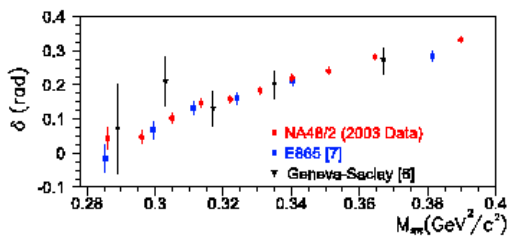}
\caption{Phase shift versus $m_\pi \pi$}  \label{phase}
\end{figure}

The Feymann diagram in Fig. \ref{ke4fey} shows one of the possibile rescattering process 
which may occur. The scattering lengths are enter in the coupling constant of the four pion vertex. 

The matrix element for $K_{e4}$ process is given by 
$$ 
T = \frac{G_F}{\sqrt{2}} V^*_{us} {\bar u}(p_\nu) \gamma _\mu 
(1 -\gamma _5) v(p_e) (V^\mu - A^\mu) $$

where the hadronic part is described using two axial ($F$ and $G$ ) and one vector (H)
 form factors \cite{bij}. After expanding them into partial wave and into
a Taylor series in $q^2 = m_{\pi \pi }^2/4m^2_+ -1$, 
the following parametrization is used to determine form factors from the
experimental data 
\begin{eqnarray}
& F = \left ( f_s + f^\prime _s q^2 + f^{\prime \prime} _s q^4 \right ) e^{i\delta_S}   \nonumber \\
& G = \left ( g_p + g^\prime _p q^2 + \right ) e^{i\delta_P} + f_p cos\theta _\pi e^{i\delta_P}  \nonumber \\
& H = h_p e^{i\delta_P} \nonumber
\end{eqnarray} 
The phase difference between $S$ and $P$ wave $\delta = \delta_S - \delta_P $ changes as   
the two pion invariant mass $m_{\pi \pi }$ increases in a way involving the parameters
$a_0$ and $a_2$ through a relation known as Roy equation \cite{roy}.
A determination of the scattering lengths may be extracted from several measurement of the phase $\delta$ at different
$m_{\pi \pi }$ values.
A measurement of $\delta$ for a particular value of  $m_{\pi \pi }$ can be done  
by measuring the $\Phi$ distibution asymmetry, where $\Phi$ is the angle between the
plane containing pions  and the one containing leptons.   
The sensitivity on $\delta$ is higher at high value of  $m_{\pi \pi }$.

Experiment NA48 collected a sample of $6.7 \times 10^5$  fully reconstructed 
$ K_{e4} $ decays. Assuming the $\Delta S = \Delta Q$ rule, which is true in the 
weak sector of the Standard Model, one can estimate the background to the $ K_{e4} $ reconstruction 
procedure by measuring  the fraction of event with ``wrong sign lepton'' or, equivalently, event with same charge
pions. The background is known to within a precision of $0.1\%$. Experimental data are summarized in  Fig. \ref{phase}. 
Let's remark that so far minimal theoretical inputs have been required and Fig. \ref{phase} is almost model independent.
On the contrary, theoretical guidance is necessary when one tries to extract the scattering lengths 
from the phase shift. The Roy equation has been implemented in the fitting as well as a theoretical constraint
between $a_0$ and $a_2$ known as universal band \cite {ub}. Additional isospin breaking effects 
have also been taken into account following indications given in \cite{iso}. 
The final results found from $K_{e4}$ analysis \cite{brigitte} are consistent with the ones found from $K \rightarrow 3\pi$
\begin{eqnarray} 
a_0 m_+ = 0.233 \pm 0.016(stat) \pm 0.007 (syst) \nonumber \\
a_2 m_+ = -0.0471 \pm 0.011(stat) \pm 0.004(syst) \nonumber
\end{eqnarray}

\section {Conclusion} 
Large statistics sample of Kaon semi-leptonic and non-leptonic decays collected by NA48 experiment 
provided precise measurements of the pion S-wave scattering lengths. Results from the two different processes
considered are consistent with each other and both agree with the result published by DIRAC collaboration \cite{Dirac} which uses a 
total different experimental approach not involving kaons. 

Results are in excellent agreement with theoretical predictions \cite{colangelo} , \cite{peleaz} and 
the level of experimental uncertainties allows one to conclude that scattering lengths measurements are so far
the most stringent test of the Chiral Theory.  The importance of these tests can be easily understood 
considering the role played by chiral symmetry effective theory in the context of flavour physics.

\begin{figure}[h]
\centering
\includegraphics[width=60mm]{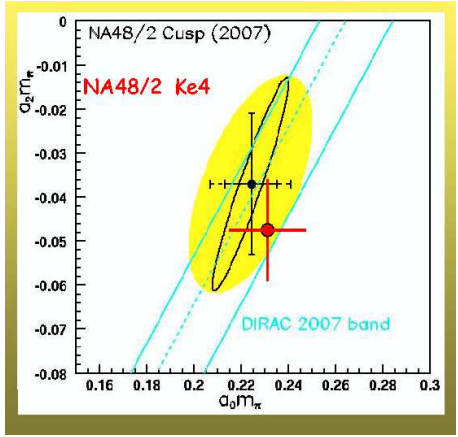}
\caption{Plane $(a_0, a_2)$, NA48 and DIRAC results comparison. } \label{concl}
\end{figure}

\section{Acknowledgments}
We gratefully acknowledge the organizers of the ``Heavy Quarks  and Leptons'' 2008 Australian edition 
and in particular Elisabetta Barberio for her wise comments about the significance 
of the measurements presented here.

\bigskip 

\end{document}